\pdfoutput=1

\documentclass[12pt]{article}
\usepackage{jheppub}

\usepackage{amsmath}
\usepackage{amssymb}
\usepackage{graphicx,color,slashed,braket}
\usepackage{bbm}
\usepackage{ifpdf}
\usepackage{comment}

\newcommand{\be}{\begin{equation}}
\newcommand{\ee}{\end{equation}}
\newcommand{\ba}{\begin{eqnarray}}
\newcommand{\ea}{\end{eqnarray}}

\newcommand{\lp}{\left(}
\newcommand{\rp}{\right)}

\newcommand{\w}{\wedge}

\def\rmi{{\rm i}}

\title{Type IIB flux compactifications with $h^{1,1}=0$}

\author{Jacob Bardzell,}
\author{Eduardo Gonzalo,}
\author{Muthusamy Rajaguru,}
\author{Danielle Smith,}
\author{and Timm Wrase}

\affiliation{Department of Physics, Lehigh University, 16 Memorial Drive East, Bethlehem, PA 18018, USA\\}

\emailAdd{jdb319@lehigh.edu}
\emailAdd{eduardo.gonzalo@lehigh.edu}
\emailAdd{muthusamy.rajaguru@lehigh.edu}
\emailAdd{des219@lehigh.edu}
\emailAdd{timm.wrase@lehigh.edu}

\notoc

\abstract{

\noindent
We revisit flux compactifications of type IIB string theory on `spaces' dual to rigid Calabi-Yau manifolds. This rather unexplored part of the string landscapes harbors many interesting four-dimensional solutions, namely supersymmetric $\mathcal{N}=1$ Minkowski vacua without flat direction and infinite families of AdS vacua, some potentially with unrestricted rank for the gauge group. We also comment on the existence of metastable dS solutions in this setup. We discuss how these solutions fit into the web of swampland conjectures. 
}

\begin{document}

\maketitle

\newpage

\section{Introduction}\label{sec:introduction}
Recent years have seen a flurry of activities related to the swampland program \cite{Palti:2019pca, vanBeest:2021lhn, Grana:2021zvf}. Many new conjectures were proposed and fascinating interconnections between different conjectures became apparent. However, given that it is extremely hard to prove any swampland conjecture (see for example \cite{Harlow:2018tng}), one might wonder whether all of the conjectures are truly imposed by quantum gravity or whether some arose from our somewhat limited understanding of string theory. Given that many conjectures are motivated and tested against what we observe in controlled string theory setups, there is an apparent need to broaden our tool kits and to get trustworthy results from larger and larger classes of string theory setups.

While this is a noble goal we are faced with the immediate problem that enlarging the landscape of trustworthy string theory constructions is extremely difficult. New approaches are rare (see for example \cite{Blumenhagen:2020xpq}) and in most instances, like for example in the absence of supersymmetry, one can only study a fairly limited set of string theories. While these might already hold surprises (see for example \cite{Dienes:2001se, Basile:2018irz, Cribiori:2020sct, Basile:2021mkd, Abel:2021tyt, Cribiori:2021txm}), we are clearly very far from even remotely understanding the landscape of non-supersymmetric string theories. 

Several conjectures involving moduli spaces and scalar potentials have been proposed recently in  \cite{Ooguri:2018wrx,Palti:2017elp, Gonzalo:2019gjp,Gonzalo:2020kke}. Sometimes these are influenced by what does and does not work in explicit string theory compactifications. The most studied of such string compactifications are based on geometric compactifications of 10d supergravity theories that arise as low energy limits of string theory. One may be tempted to conjecture that a property that we observe in such a corner of moduli space is indeed a fundamental consistency requirement of quantum gravity. However, the presence of a heuristic arguments, for example based on black hole physics, is usually considered a more important hint that the criteria applies more generally. In the absence of such arguments one can only try to enlarge the landscape of four-dimensional theories that one can obtain from string theory, to check existing conjectures against a larger part of the string landscape. This is what we are doing here, by focusing on a corner of the landscape which has not been explored much, namely flux compactifications with orientifolds and purely non-geometric SCFT descriptions for the internal dimensions. In particular, we focus on type IIB and two different Landau-Ginzburg (LG) orientifold models, with $F_{3}$ and $H_{3}$ fluxes turned on. The tools necessary to determine the low energy effective action for these models were spelled out in \cite{Becker:2006ks}.  

In this work we revisit and expand the results of \cite{Becker:2006ks, Becker:2007dn, Ishiguro:2021csu}. We find new AdS, Minkowski and dS solutions and discuss them in the context of the swampland program. In the first part of the paper we focus our attention on supersymmetric solutions. At weak coupling and large complex structure we find several infinite families of AdS solutions. In some cases the solutions are mirror dual to the well known type IIA AdS flux vacua found by DeWolfe, Giryavets, Kachru and Taylor (DGKT) \cite{DeWolfe:2005uu} (see also \cite{Behrndt:2004mj, Derendinger:2004jn, Lust:2004ig, Grimm:2004ua, Camara:2005dc} for earlier work). We find perfect agreement with the AdS distance conjecture except in one family of solutions which is dual to DGKT. It was argued in \cite{Buratti:2020kda} that the presence of a $\mathbb{Z}_k$ discrete symmetry forces a modification of the conjecture. We argue that such a symmetry is indeed present in our setup so we find an agreement with the refined version of the conjecture, just like for DGKT. 

Surprisingly our setup, which is essentially dual to a generalization of the DGKT model in type IIA, allows also for infinite families of AdS solutions with an ever growing number of D3 branes. Such solutions do naively give rise to AdS$_4$ spacetimes with arbitrarily large gauge group rank. We are not aware of any such solutions in the literature and they certainly deserve further study and scrutiny.

We also find fully stabilized four-dimensional Minkowski families of solutions, which are to our knowledge the only full-fledged string theory constructions of $\mathcal{N}=1$ Minkowski vacua without flat directions. Such Minkowski vacua were previously discovered in \cite{Becker:2006ks, Becker:2007dn} and their validity beyond the perturbative regime was shown to be guaranteed by a powerful non-renormalization theorem in \cite{Becker:2006ks}. We extend the previous analysis to show that these vacua do not only survive despite string loop corrections but we also prove that, although string loop corrections can change the masses, they cannot lead to any flat directions.

We also study non-supersymmetric solutions. In the parametrically weak coupling and large complex structure regime we find a family of non-supersymmetric AdS vacua as in the dual DGKT setting. We find that the masses lie above the BF bound, so the vacuum is perturbatively stable. According to the conjecture in \cite{Ooguri:2016pdq} these should be unstable but we leave it for the future to study potential decay.   
In the not-so-weak coupling regime we find a metastable dS vacuum which requires the presence of D3 branes in order to satisfy the tadpole cancellation condition. These vacua are, however, not protected by the non-renormalization theorem. In particular, the K\"ahler potential is expected to receive quantum corrections that are not under control and therefore these dS vacua are not trustworthy.

The structure of the paper is as follows. In the next section \ref{sec:review} we review how to obtain the low energy 4d $\mathcal{N}=1$ theories and revisit the no-go theorems protecting the superpotential. Then, in section \ref{sec:Mink} we discuss fully stabilized, supersymmetric Minkowski vacua and contrast their existence with related swampland conjectures. In section \ref{sec:AdS} we find several new families of supersymmetric and non-supersymmetric AdS vacua and we discuss their connection to the AdS distance conjecture. Lastly, we study dS vacua in section \ref{sec:dS} before concluding in section \ref{sec:conclusion}. We include several useful formulas regarding Type IIA flux compactifications on Calabi-Yau manifolds in appendix \ref{app:dualIIA}.

\section{Review of the setup}\label{sec:review}
In type IIA flux compactifications on Calabi-Yau manifolds with smeared O6-plane sources and NSNS and RR fluxes it is possible to stabilize all moduli at tree-level if $h^{2,1}=0$, i.e. if we are dealing with a rigid Calabi-Yau manifold \cite{DeWolfe:2005uu}. In a mirror symmetric type IIB compactification, using the SYZ conjecture \cite{Strominger:1996it}, one would then expect to be able to stabilize all moduli on `spaces' with $h^{1,1}=0$. The RR fluxes $F_p$ with $p=0,2,4,6$ on the IIA side all transform into RR $F_3$ flux. The IIA $H_3$ flux could in principal transform to a mixture of NSNS $H_3$, geometric and non-geometric fluxes in IIB. However, on the type IIA side, for $h^{2,1}=0$, we have a space with only one 3-cycle (and its dual). Turning on the $H_3$ flux in type IIA so that it does not thread the $T^3$ fibration of the SYZ setup, we expect that after the three T-dualities, we end up in type IIB with a setup that only involves $H_3$ flux and neither geometric nor non-geometric fluxes. Intuitively, this might also be expected from the work of Giddings, Kachru and Polchinski (GKP) \cite{Giddings:2001yu} that showed that in type IIB it is possible to stabilize the axio-dilaton and all complex structure moduli using only $F_3$ and $H_3$ fluxes. This means of course that we can stabilize all moduli in the absence of K\"ahler moduli, i.e. for $h^{1,1}=0$. 

This idea of studying how all moduli are stabilized at tree-level in type IIB flux compactifications with $F_3$ and $H_3$ fluxes on `spaces' with $h^{1,1}=0$ was first fleshed out in \cite{Becker:2006ks, Becker:2007dn}. There the authors studied orbifolds of certain Landau-Ginzburg models and searched successfully for completely stabilized, supersymmetric 4d $\mathcal{N}=1$ Minkowski and AdS vacua. Such Minkowski vacua are absent in geometric type IIA flux compactifications \cite{Micu:2007rd, Ihl:2007ah} and require non-geometric fluxes, which are not well-controlled due to potential $\alpha'$ corrections. However, as mentioned above, under mirror symmetry the $H_3$ flux can become geometric and non-geometric fluxes. So, even if we only turn on the $H_3$ flux on the type IIB side, we actually probe a genuinely larger part of the string landscape than DGKT. Furthermore, due to powerful no-go theorems that we will review in the next subsection, these settings are very well-controlled. 

Landau-Ginzburg orbifold models provide a way of analytically continuing Calabi-Yau compactifications to small volume and can even be used to describe the mirror dual of compactification on a rigid Calabi-Yau manifold \cite{Hori:2000kt}. A Landau-Ginzburg theory is determined by the superpotential $W(\Phi_i)$, which is a quasi-homogeneous analytic function of the (worldsheet) chiral superfields $\Phi_i$. In this paper, following \cite{Becker:2006ks}, we will consider two models. Firstly, we consider the $1^9$ model with a superpotential given by
\begin{equation}
    W = \sum_{i=1}^{9} \Phi_i^3\,,
\end{equation}
and secondly we will consider the $2^6$ model with a superpotential given by
\begin{equation}
    W = \sum_{i=1}^{6} \Phi_i^4.
\end{equation}
In the $1^9$ model one can orbifold by the $\mathbb{Z}_3$ symmetry $\Phi_i \rightarrow \omega \Phi_i$ where $\omega = e^{\frac{2 \pi \rmi}{3}}$, while in the $2^6$ model we use the $\mathbb{Z}_4$ symmetry with $\omega = e^{\frac{\pi \rmi }{2}}$. For the $1^9$ orientifold, $\sigma_1$ in \cite{Becker:2006ks}, one combines worldsheet parity with $(\Phi_1,\Phi_2,\Phi_3,...,\Phi_9) \rightarrow -(\Phi_2,\Phi_1,\Phi_3,...,\Phi_9)$. The orientifold for the $2^6$ model is the $\sigma_0$ orientifold in \cite{Becker:2006ks} that acts on the fields as $(\Phi_1,\Phi_2,\Phi_3,...,\Phi_6) \rightarrow e^{2\pi \rmi/8} (\Phi_1,\Phi_2,\Phi_3,...,\Phi_6)$. In both of the cases one ends up with O3-planes whose charge can be cancelled by turning on $F_3$ and $H_3$ fluxes and/or by adding D3 branes.

Before turning on the fluxes, it is easy to check which are the corresponding Calabi-Yau (CY) manifolds. We need to compute the dimensions of the ring of superprimary chiral operators $R=\frac{C[\Phi]}{\partial_{j}W(\Phi)]}$. The $(c,c)$ ring correspond to $(2,1)$ harmonic forms while the chiral-antichiral ring $(c,a)$ corresponds to $(1,1)$ forms. For the $1^9$ model it is easy to check that there are $h_{2,1}=63$ monomials $\Phi_{i}\Phi_{j}\Phi_{k}$ which are invariant under the $\mathbb{Z}_3$ and the orientifold action. One also obtains $h_{1,1}=0$ \cite{Becker:2006ks}, that is, there are no corresponding K\"ahler moduli in the would be CY manifold. It corresponds to the mirror of $\frac{T^6}{\mathbb{Z}_3 \times \mathbb{Z}_3}$. Thus we see that in the absence of fluxes the model is dual to a DGKT construction, i.e. to a compactification of type IIA on a rigid CY manifold.\footnote{The actual model that was explicitly worked out by DGKT is a slightly different $\frac{T^6}{\mathbb{Z}_3 \times \mathbb{Z}_3}$ that differs from this model in the twisted sector \cite{Becker:2007dn}.} Similarly, for the $2^6$ orientifold one obtains $h^{1,1}=0$ and $h_{2,1}=90$ and it corresponds to the mirror of $\frac{T^6}{\mathbb{Z}_4 \times \mathbb{Z}_4}$. 

In this work, following \cite{Becker:2007dn, Ishiguro:2021csu}, we will restrict ourselves to what would be the bulk moduli in the mirror dual toroidal orbifold. We will furthermore set the three bulk complex structure moduli equal and study a rather simple 4d $\mathcal{N}=1$ $SU$ model. This allows us to find many analytic families of solutions and thereby truly study the parameter space of this model in great detail.\footnote{While the previous work \cite{Becker:2006ks, Becker:2007dn} studied particular solutions of these models, the more recent paper \cite{Ishiguro:2021csu} picked random flux numbers within a finite range and generated large \emph{generic} solution sets that were compared with a variety of swampland conjectures. Here, we actually test several swampland conjectures against new infinite families of analytic solutions.} It is expected that all of our findings carry over to the full model. In the simplest, somewhat restricted setup where our model is dual to the DGKT model, this follows from the explicit check of the blow-up modes in the DGKT paper \cite{DeWolfe:2005uu}. When talking about fully stabilized Minkowski vacua, then we can refer to the paper \cite{Becker:2006ks} where Minkowski vacua were found even when including all moduli. In particular, our new proof below that the masses cannot become zero even when including all corrections applies equally well to our $SU$ model and the full model studied in \cite{Becker:2006ks}. However, although we do not expect surprises, it would of course be interesting to extend our analysis to a generic setup with arbitrary many moduli.

The careful reader might worry that stabilizing blow-up modes requires turning on many additional fluxes that then contribute to the tadpole which then might become much larger than the fixed negative charge induced by the O3 planes in our models. This expectation would be in line with the recently proposed tadpole conjecture \cite{Bena:2020xrh, Bena:2021wyr, Plauschinn:2021hkp, Lust:2021xds, Bena:2021qty}. However, it does not apply here for two reasons: Firstly, in the case where our models are dual to the DGKT model, all blow-up moduli are stabilized in the dual model by using $F_4$ fluxes that do not appear in any non-trivial tadpole condition in the type IIA model. This means that the dual $F_3$ flux quanta, that stabilize blow-up modes, likewise do not appear in the tadpole condition on the type IIB side. Secondly, as we explain in the next subsection, the large volume intuition that fluxes contribute with the same sign as D3 branes to the tadpole is not correct in these non-geometric settings. Fluxes are no longer required to be ISD and can even in supersymmetric solutions contribute to the tadpole with the same sign as orientifold planes.

Type IIB string theory compactifications on the above two Landau Ginzburg models, after including the above discussed O3 orientifold projections, give rise to 4d $\mathcal{N}=1$ theories. The superpotential is generated by $H_3$ and $F_3$ fluxes and takes the standard form $W=\int_M (F_3-S H_3) \w \Omega$ \cite{Dasgupta:1999ss, Gukov:1999ya}. However, given that we are in these setting in a small volume regime, the usual K\"ahler potential $K=-\ln[-\rmi(S-\bar S)] - \ln[\rmi \int_M \Omega \wedge \overline{\Omega}]$ does receive corrections as discussed in subsection 3.2 of \cite{Becker:2007dn}. These corrections can be derived by using mirror symmetry (see appendix \ref{app:dualIIA}), which leads to the following the K\"ahler potential $K=-{\mathbf 4} \ln[-\rmi(S-\bar S)] - \ln[\rmi \int_M \Omega \wedge \overline{\Omega}]$. In our simple case where we restrict to two moduli, the axio-dilaton $S=C_0+\rmi e^{-\phi}$ and a complex structure modulus $U$, both the $1^9$ and the $2^6$ model give rise to the following K\"ahler and superpotential
\ba
K &=&-4 \ln [-\rmi(S-\bar{S})]-3 \ln [-\rmi(U-\bar{U})]\,, \label{eq:K}\\
W &=&\left(f^{0}-S h^{0}\right) U^{3}-3\left(f^{1}-S h^{1}\right) U^{2}+3\left(f_{1}-S h_{1}\right) U+\left(f_{0}-S h_{0}\right)\,.\label{eq:W}
\ea
This restricted model, corresponding to $h^{2,1}=1$, is dual to a similarly restricted model in type IIA where, for example, one sets the three K\"ahler moduli in the original DGKT model equal, to get an effective model with $h^{1,1}=1$ on the type IIA side (an $ST$ model). The four $F_3$ flux components, labelled by $f_0,f_1,f^1,f^0$ above correspond on the type IIA side to $F_6$, $F_4$, $F_2$ and $F_0$ fluxes, while the four $H_3$ flux components $h_0,h_1,h^1,h^0$ correspond on the type IIA side to $H_3$ flux, metric flux and non-geometric $Q$ and $R$ fluxes, respectively (see table 1 in \cite{Shelton:2005cf}). Thus, this flux compactification on the type IIB side is indeed extending the original DGKT construction \cite{DeWolfe:2005uu} in a very important way. Furthermore, as we will explain in the next subsection, there are non-renormalization theorems that allows one to obtain trustworthy results in regimes that have not really been probed much in the existing literature.

As is familiar from any flux compactification with orientifolds, one has to cancel the net charge induced by the fluxes, O-planes and potentially D-branes. In our case this will be the O3 plane charge and the tadpole condition is given by
\be
\int_M F_3 \w H_3 + N_{D3} = \frac12 N_{O3}\,.\label{eq:tadpole}
\ee
This allows us now to clarify, why we discussed above the $1^9$ model and the $2^6$ model although they both give rise to the same (restricted) K\"ahler and superpotential in equations \eqref{eq:K} and \eqref{eq:W}: The above mentioned orientifold projection for the $1^9$ model gives rise to $N_{O3}=24$, while the orientifold projection for the $2^6$ model gives rise to $N_{O3}=80$ \cite{Becker:2006ks}. This means that the flux contribution to the tadpole
\begin{equation}
N_{\text{flux}} = \int_M F_3 \w H_3 = -h^{0} f_{0} - 3 h^{1} f_{1} + h_{0} f^{0} + 3 h_{1} f^{1} \,,
\end{equation}
would have to equal either 12 or 40, if we want to satisfy the tadpole condition in equation \eqref{eq:tadpole} without the addition of D3 branes.

However, it is also important and interesting in these models to include D3 branes. The reason for this is that the flux contribution to the tadpole $N_{\text{flux}}$ has no definite sign (see subsection 3.3 in \cite{Becker:2007dn}). This means in particular that fluxes can contribute with the same sign as O3 planes in the tadpole and we will see below that there are even infinite families in which $N_{\text{flux}} \rightarrow -\infty$ and at the same time $N_{D3} \to \infty$. One may ask why this is possible, since in the well-known geometric type IIB CY orientifolds with 3-form fluxes, studied in GKP \cite{Giddings:2001yu}, the $N_{\text{flux}}$ is always positive. This follows in that case simply from the requirement that the flux $F_3 - S H_3$ has to be imaginary self dual (ISD). The latter in turn follows from the vanishing of the covariant derivatives of the superpotential with respect to the axio-dilaton and the complex structure modulus, i.e. $D_SW=D_UW=0$. In our setup there are small volume corrections to the K\"ahler potential in equation \eqref{eq:K}. In particular, the factor of 4 changes $D_SW=0$ in such a way that one can no longer derive the ISD requirement, as is discussed in more detail in \cite{Becker:2007dn}. 

The above property might be surprising.\footnote{At small volume there are a plethora of instances were the large volume understanding of mutually supersymmetric objects changes completely due to stringy corrections. So, it shouldn't necessarily be surprising that fluxes can carry anti-D3 brane charge and still be mutually supersymmetric with D3 branes and O3 planes.} Therefore we quickly discuss it also in the dual type IIA models. In the case where we only turn on a single $H_3$ flux quanta our model is dual to a type IIA flux compactification à la DGKT with $h^{2,1}=0$. There is then a single O6 plane tadpole condition. In this case, for supersymmetric AdS vacua, the flux contribution to the tadpole has to have the same sign as D6-branes. If that were not the case, then we could add D6 branes in addition to the O6 planes and thereby completely cancel their negative contribution to the scalar potential. Then there would be no negative term in the scalar potential (and therefore no AdS vacua) since RR fluxes and $H_3$ flux contribute positive definite terms only. Thus in this case, which is the dual of the DGKT model with $h^{2,1}=0$, the fluxes induce a charge in the tadpole that has the same sign as D branes and $N_{\text{flux}}$ is therefore bounded by zero from below and $N_{O3}/2$ from above. Now when we turn on more general $H_3$ flux on the type IIB side then this corresponds to type IIA flux compactifications in the presence of geometric and non-geometric fluxes. These fluxes can contribute to the scalar potential with either sign and the O plane term is no longer the only negative term in the scalar potential. Thus, there is no immediate obstruction to over-cancelling the O plane contribution by adding a very large number of D branes. We will see how this works in explicit examples below, when we discuss concrete solutions.

\subsection{Non-renormalization theorem}
In this subsection we first recall the absence of perturbative and non-perturbative corrections to the superpotential \cite{Becker:2006ks, Becker:2007dn}. First of all, $\alpha^{\prime}$ corrections are already taken into account in the LG theory. Thus, one only has to focus on $g_s$ perturbative and non-perturbative corrections. However, it was argued in \cite{Becker:2006ks,Becker:2007dn} that the superpotential does not receive any perturbative or non-perturbative corrections at all, which follows for example from the non-renormalization of the BPS tension of a D5-brane domain wall but also passes other non-trivial checks \cite{Becker:2006ks}. This means that the superpotential is exact even at strong coupling. Note however, that the K\"ahler potential can and will receive perturbative and non-perturbative corrections, which is something we will return to in the next paragraph. The cautious reader might worry about the familiar brane instanton corrections to the IIB superpotential. Let us therefore recall that our models have $h^{1,1}=0$ and thus no Euclidean D3-brane instantons. The absence of D(-1) instantons was argued for in footnote 6 in \cite{Becker:2006ks} as follows: Since the D(-1) instantons do not depend on the volume and they are not there in the decompactification limit due to higher supersymmetry, they should also not appear here. This is also consistent with the recent analysis in \cite{Kim:2022jvv}, which trivially covers our setup since we have $h^{1,1}=0$ and therefore no 4-cycles and no D7 branes or O7 planes. Alternative it was argued for the absence of any brane instanton corrections in \cite{Becker:2007dn} using the duality to the type IIA setting of DGKT: There the only 3-cycle in models with $h^{2,1}=0$ has $H_3$ flux and therefore there are no brane instantons \cite{Marino:1999af}.

When studying Minkowski vacua we will assume that the non-renormalization theorem holds and the superpotential receives no corrections even in those vacua where $g_s$ is of order 1 or larger. The conditions for supersymmetric Minkowski vacua are $\partial_i W = W = 0$ and do not depend on the K\"ahler potential. Thus, the very existence of Minkowski vacua does not change even if one includes arbitrary corrections because those can only appear in the K\"ahler potential. Previously, such explicit supersymmetric, fully stabilized Minkowski vacua where constructed in \cite{Becker:2006ks, Becker:2007dn, Ishiguro:2021csu}. However, it was stated in \cite{Becker:2007dn} that these are necessarily at strong coupling\footnote{We find that they cannot be at parametrically weak coupling but there are certainly examples with $g_s <1$.} and thus receive large corrections to the K\"ahler potential. This then leads to the following important question: Are these truly fully stabilized 4d $\mathcal{N}=1$ Minkowski vacua or can the corrections to $K$ lead to flat directions?

We will prove here that even arbitrary, unknown corrections to $K$ cannot lead to flat directions in these models: We assume that one has been able to a find a fully stabilized SUSY Minkowski vacuum as was the case in \cite {Becker:2006ks, Becker:2007dn, Ishiguro:2021csu} (see also section \ref{sec:Mink} below). Then the Hessian matrix of second derivatives of the scalar potential has only positive eigenvalues and is given by\footnote{For simplicity we work here with the Hessian. The actual masses squared are the eigenvalues of $H_{i\bar{\jmath}} K^{\bar \jmath k}$. However, given that the K\"ahler metric is positive definite, this does not change our conclusion.}
\begin{equation}
H_{i\bar{\jmath}}=\partial_{i}\partial_{\bar{\jmath}}V=(\partial_{i}\partial_{k}W)K^{k\bar{\ell}}(\partial_{\bar \ell}\partial_{\bar \jmath}\overline{W}),
\end{equation}
or in matrix form
\begin{equation}
H=\mathcal{W}\mathcal{K}\overline{\mathcal{W}}\,.
\end{equation}
Now compute the determinant
\begin{equation}
\det\,H=\det\,\mathcal{W}\,\det\,\mathcal{K}\,\det\,\overline{\mathcal{W}}=|\det\,\mathcal{W}\,|^{2}\det\,\mathcal{K}\,.
\end{equation}
Given that \emph{all} eigenvalues of $H$ were positive to begin with we can conclude that $|\det\,\mathcal{W}\,|^{2}>0$.

Now let us take into account arbitrary and unknown corrections to the K\"ahler potential and denote the inverse K\"ahler metric after including all these corrections $\mathcal{K}_{c}$. The new Hessian for this corrected Minkowksi vacuum is now given by
\begin{equation}
H_{c}=\mathcal{W}\mathcal{K}_{c}\overline{\mathcal{W}}\,.
\end{equation}
We again compute the determinant
\begin{equation}
\det\,H_{c}=\det\,\mathcal{W}\,\det\,\mathcal{K}_c\,\det\,\overline{\mathcal{W}}=|\det\,\mathcal{W}\,|^{2}\det\,\mathcal{K}_c\,.
\end{equation}
Since the superpotential did not receive any corrections we have from above that $|\det\,\mathcal{W}\,|^{2}>0$. Since the K\"ahler metric controls the kinetic terms of the scalar fields, its eigenvalues have to be positive. This remains true even after including arbitrary corrections and therefore $\det\,\mathcal{K}_c\neq0$. This, combined with the preservation of $|\det\mathcal{W}|^{2}$ implies that $\det\,H_{c}\neq0$. Thus, \emph{all} the eigenvalues of $H_{c}$ must be nonzero. 

In supersymmetric Minkowski vacua eigenvalues of the Hessian matrix have to be positive for stabilized moduli or zero for flat directions. It was just shown that the eigenvalues of $H_{c}$ are nonzero, so we can conclude that these Minkowski vacua cannot have flat directions even when including unknown and arbitrary corrections to the K\"ahler potential.

One can actually prove also the existence of AdS vacua at strong coupling using the non-renormalization of $W$ \cite{Becker:2006ks}. While this is not so important since there are infinite families of AdS vacua with parametrically weak coupling, let us nevertheless briefly recall the argument. For supersymmetric AdS vacua, satisfying $D_i W=\partial_i W+(\partial_i K) W=0$, the $\partial_{i}K$ term can receive corrections. The authors of \cite{Becker:2006ks} expanded the corrected K\"ahler potential around the minimum which one can choose to be at $\phi^i=0$, so that $K_c = K+f(\phi^i) +\bar{f}(\bar{\phi}^{\bar{\imath}}) +\phi^i\bar{\phi}^{\bar{\jmath}} g_{i\bar{\jmath}}(\phi^i,\bar{\phi}^{\bar{\imath}})$. At the minimum $\phi^i=0$ the only correction to $\partial_{i}K$ arises from $f(\phi^i)$. However, this can be interpreted as a K\"ahler transformation: $K\to K+f+\bar f$, $W\to W e^{-f}$, which changes $D_iW \to e^{-f} D_iW$. Therefore, supersymmetric AdS vacua satisfying $D_iW=0$ cannot disappear even when including arbitrary unknown K\"ahler corrections. However, for example the mass spectrum is expected to be corrected (within the limits allowed by $\mathcal{N}=1$ supergravity).

Finally, there is no argument for preventing corrections to non-supersymmetric vacua. So, if one finds them at strong coupling, they could disappear or become unstable when including string loop corrections. Here however let us recall that the Landau-Ginzburg model takes all $\alpha'$ corrections into account, so we do not need to be in the large complex structure limit to trust these solutions \cite{Becker:2006ks}.

\section{Fully stabilized $\mathbf{\mathcal{N}=1}$ Minkowski vacua}\label{sec:Mink}
As mentioned previously, the first fully stabilized 4d $\mathcal{N}=1$ Minkowski vacua were found in \cite{Becker:2006ks}. In the dual type IIA case, such vacua do not exist in geometric compactifications \cite{Micu:2007rd, Ihl:2007ah}, which means that in the type IIB models at least two components of the $H_3$ flux have to be turned on. It was also show in \cite{Becker:2007dn} that these IIB Minkowski vacua are never arising at large complex structure, i.e. on the dual type IIA side they cannot arise at large volume. However, as we reviewed above the Landau-Ginzburg models take all $\alpha'$ corrections into account and therefore do not require us to be at large complex structure. It was furthermore stated \cite{Becker:2007dn} that these Minkowski vacua are confined to strong coupling. Given the non-renormalization theorem from the previous section, we can trust Minkowski vacua even at strong coupling. However, we also find that only parametrically weak coupled solutions are forbidden in this setup and $g_s<1$ is possible with a model dependent lower bound on $g_s$. In the next subsection we present a new infinite family of fully stabilized supersymmetric Minkowski vacua and in the following subsection we discuss how this family of solutions fits into the swampland program.

\subsection{Minkowski solutions}
In order to find Minkowski vacua we have to solve $W=\partial_S W = \partial_U W=0$ for the $W$ given in equation \eqref{eq:W} above. A particular family of solutions with properly integer quantized fluxes arises for 
\be
f^{0} = -4, \quad f^{1} = 0,\quad f_{1} = 0,\quad f_{0} = 4,\quad h^{0} = -3 - h_{0},\quad h^{1} = 1,\quad h_{1} = -1\,.
\ee
Here $h_0 \in \mathbb{Z}$ is a free parameter that actually does not appear in the tadpole condition since $N_{\text{flux}}$ in equation \eqref{eq:tadpole} reduces to $N_{\text{flux}}=12$ independent of $h_0$. Thus, this is a solution to the $1^9$ model which does not require D3 branes since the fluxes cancel the negative O3 plane charge.

The moduli are stabilized at the following values
\ba
&\text{Re}(U) = -\frac12\,,\qquad\qquad &\text{Im}(U) = \frac{\sqrt{3}}{2}\,,\cr
& \text{Re}(S) = \frac{6 + 4 h_{0}}{3 + h_{0} (3 + h_{0})}\,, \qquad \qquad &\text{Im}(S) = \frac{2 \sqrt{3}}{3 + h_{0} (3 + h_{0})}\,.
\ea
While the complex structure modulus is stabilized at a fixed value, the inverse string coupling Im$(S)$ changes when we vary the free parameter $h_0 \in \mathbb{Z}$. It takes on its maximal value of Im$(S)=2\sqrt{3}$ for $h_0=-1$ and for $h_0=-2$. For $h_0 \to \pm \infty$ we enter parametrically strong coupled regions with Im$(S) \propto 1/h_0^2$. We stress again that even in this parametrically strong coupled regime there are no corrections to $W$ due to the above non-renormalization theorem.

The positive masses squared for the two complex scalar fields in the Minkowski vacuum are given by
\be
m_-^2 = \frac{(11 - 4 \sqrt{7}) (3 + h_0 (3 + h_0))^3}{192 \sqrt{3}}\,,\qquad m_+^2 = \frac{(11 + 4 \sqrt{7})(3 + h_0 (3 + h_0))^3}{192 \sqrt{3}}\,.
\ee
We see that in the limit $h_0 \to \pm \infty$ the masses grow like $h_0^6$. For the largest inverse string coupling value Im$(S)=2\sqrt{3}\approx 3.46$ which is obtained for $h_0=-1$ and for $h_0=-2$, the masses squared reduce in both cases to $m_-^2 = \frac{11 - 4 \sqrt{7}}{192 \sqrt{3}}\approx 0.00125$ and $m_+^2 = \frac{11 + 4 \sqrt{7}}{192 \sqrt{3}}\approx 0.0649$.

\subsection{Minkowski vacua and the swampland}
It is easy to find string compactifications that give rise to 4d Minkowski vacua with $\mathcal{N} \geq 2$, for example, by compactifying type II string theory on a Calabi-Yau manifold or a torus. However, to the best of our knowledge all these Minkowksi vacua with $\mathcal{N}\geq 2$ have flat directions, i.e. massless scalar fields. These flat directions can be protected by the high amount of supersymmetry. However, in 4d $\mathcal{N}=1$ theories there is no such protection and it is expected that all flat direction would be lifted by corrections which likely leads to runaway directions. To the best of our knowledge, the Minkowski vacua first discovered in \cite{Becker:2006ks, Becker:2007dn} are the only fully stabilized $\mathcal{N}=1$ Minkowski vacua that arise in full-fledged string theory constructions. Given that corrections to the scalar potential are not forbidden by $\mathcal{N}=1$ supersymmetry, one would have thought that it would not be possible to really argue for the existence of these vacua when including all perturbative and non-perturbative corrections. However, the non-renormalization of the superpotential \cite{Becker:2006ks} and our argument above about the mass matrix are implying that these vacua do indeed exist in a strongly coupled corner of string theory. 

Given the more recent objections to the existence of dS vacua in string theory \cite{Danielsson:2018ztv, Obied:2018sgi}, the very existence of fully stabilized 4d $\mathcal{N}=1$ Minkowksi vacua was questioned as well. The reason is that any small, SUSY breaking, positive energy contribution to the scalar potential turns these Minkowksi vacua into metastable dS solutions. Following this logic, the authors of \cite{Gautason:2018gln} conjectured that strongly stabilized AdS vacua should be forbidden. Here by strongly stabilized one means that the mass of the lightest field satisfies $m_{\text{light}} L_{AdS} \gg 1$, where $L_{AdS}$ is the length scale of the AdS space. This AdS moduli conjecture seems to imply that if we take the limit $L_{AdS} \to \infty$ to go to Minkowksi space, then $m_{\text{light}} \to 0$ in contradiction with the Minkowski vacua discussed here and previously in \cite{Becker:2006ks,Becker:2007dn,Ishiguro:2021csu}. Note however, that these Minkowski vacua cannot arise as the limit of any of the infinite families of AdS solutions that we find in these models and that will be discussed in the next subsection. Likewise, there is no obvious small SUSY breaking correction or change to the model that leads to dS vacua. All string loop corrections do not change $W$ and only modify the values of the positive masses squared of the scalar fields in the Minkowski vacuum. Changing some flux quanta to break supersymmetry is a large effect and the same probably applies to any other change given that the complex structure modulus is stabilized at order 1 and we are at strong coupling. However, it would definitely be interesting to study this further.

The existence of these vacua and the absence of corrections is surprising, maybe even more so given the recent paper \cite{Palti:2020qlc} that finds that generically in quantum gravity any allowed correction should appear. The exception to this rule is stated in the same paper and is formalized in the supersymmetric genericity conjecture \cite{Palti:2020qlc}. This conjecture says that quantities that are protected in higher supersymmetric theories should only vanish in lower supersymmetric theories, if the lower supersymmetric theory is related to a higher supersymmetric theory. In particular, the authors discuss 4d $\mathcal{N}=1$ Minkowski vacua with everywhere vanishing superpotential, $W=0$. They find that the equation $W=0$ can only survive all corrections if the theory is related to a higher dimensional theory via for example a simple orbifold projection. While our setup with fluxes and a non-zero $W$ generate by those fluxes is not covered by the analysis in \cite{Palti:2020qlc}, our findings seem nevertheless compatible with the supersymmetric genericity conjecture since our setups are simple orbifolds of toroidal models that preserve higher amounts of supersymmetry.

Summarizing, these non-geometric type IIB setups give rise to fully stabilized 4d $\mathcal{N}=1$ Minkowksi vacua that seem to survive all stringy corrections, which makes them to our knowledge the only full-fledged string theory constructions of this type. These vacua arise only at relatively strong coupling in a barely studied part of the string landscape.

\section{Infinite families of AdS vacua}\label{sec:AdS}
In this section we study exemplary families of AdS solutions that arise in these non-geometric type IIB flux compactifications. As discussed above, due to the non-renorma-lization of $W$ even supersymmetric AdS solutions at strong coupling will persist when including all potential corrections. However, for example the masses and the cosmological constant in these solutions might get significantly modified when we are not at weak string coupling. All the different families of AdS solutions that we present below, allow us to go to parametrically weak coupling and thus we have parametric control over them. This enables us to perform trustworthy and detailed studies even when these solutions are not supersymmetric. Given that the exact number of O3 planes in these infinite families plays essentially no role, we will restrict ourselves to the $1^9$ model with $N_{O3}=24$. We will introduce representative examples to illustrate the different behaviors that these infinite families display. Firstly, we present families that are dual to the AdS vacua found in DGKT \cite{DeWolfe:2005uu} but we also find other infinite families of AdS vacua that arise in our more general setup. Secondly, we study interesting and very different sets of solutions, where by increasing the number of D3 branes the contribution of the fluxes to the tadpole can become negative and very large. In the $N_{\text{flux}}\rightarrow -\infty$ limit the number of D3 branes needs to become infinite $N_{D3} \to \infty$ as well, in order to satisfy the tadpole condition. We discuss how all these solutions fit into the web of swampland conjectures at the end of this section.

\subsection{Infinite families of AdS vacua without D3 branes}\label{ssec:AdSnoD3}

\subsubsection{The DGKT dual}\label{ssec:DGKTAdS}
In \cite{Becker:2007dn} two infinite families of SUSY AdS solutions were presented. The first solution is related to the infinite family of SUSY AdS vacua that were found in DGKT \cite{DeWolfe:2005uu}.\footnote{In the second SUSY AdS solution in subsection 4.3.2 in \cite{Becker:2007dn}, there seems to be a typo. We find that either Im$(U)$ or Im$(S)$ are necessarily negative, so this does not seem to be a physically meaningful solution.} To find it one has to necessarily set three $H_3$ flux quanta to zero, $h^{0}=h^{1}=h_{1}=0$. The tadpole condition \eqref{eq:tadpole} then implies
\begin{equation}\label{eq:tadpoleDGKTAdS}
h_{0} = \frac{12}{f^{0}}\,,
\end{equation}
which means that due to flux quantization $f^0 \in \{1,2,3,4,6,12\}$. We will not plug in any specific flux values but keep in mind that $f^0$ and $h_0$ are bounded due to tadpole cancellation condition but the other fluxes are not.

One can easily solve $D_S W=D_UW=0$ and find that the axio-dilaton is stabilized at 
\begin{equation}
\text{Re}(S) = \frac{f_{0} (f^{0})^2 + 3 f_{1} f^{0} f^{1} - 2 (f^{1})^3}{12 f^{0}}\,, \qquad  \text{Im}(S) = 2 \sqrt{\frac{5}{3}}\frac{\left( f_{1} f^{0}-(f^{1})^{2}\right)^{\frac32}}{9 f^{0}}\,,
\end{equation}
while the complex structure modulus is stabilized at 
\begin{equation}
\text{Re}(U) =  \frac{f^{1}}{f^{0}}\,, \qquad \text{Im}(U) =\sqrt{\frac{5}{3}} \frac{ \lp f_{1} f^{0}-(f^{1})^{2}\rp^{\frac12}}{ f^{0}}\,.
\end{equation}
Given that $f_1$ is unconstrained by the tadpole, we can make it large and even send it to infinity. In that limit the string coupling $1/\text{Im}(S)$ becomes parametrically small and the complex structure modulus becomes parametrically large. This is the mirror dual of the large volume, weak coupling families of AdS vacua that arise in type IIA flux compactifications if one makes the $F_4$ flux large \cite{DeWolfe:2005uu}.

The scalar potential at the minimum is 
\begin{equation}
    V_{AdS} = \frac{- 19683 \sqrt{\frac{3}{5}} (f^{0})^{3}}{3200 (f_{1} f^{0} -  (f^{1})^{2})^{\frac92}}\,.
\end{equation}
The four masses squared in this family can be conveniently expressed in terms of the above value of the scalar potential as
\begin{equation}
   m^2 =  \left\{\frac{10}{3}, 6, \frac{70}{3}, \frac{88}{3}\right\} |V_{AdS}|\,.
\end{equation}
Since the AdS radius in 4d is given by $R_{AdS}=\sqrt{3/|V_{AdS}|}$ one finds surprisingly that all the masses squared in AdS units, i.e. all $m^2 R_{AdS}^2$, are integers. This was recently discovered in \cite{Conlon:2021cjk} (see also \cite{Marchesano:2020uqz}). Furthermore, the integers are such that the operator scaling dimensions in the dual CFT$_3$, i.e. 
\be
\Delta=\frac12\lp 3 +\sqrt{9+4m^2 R_{AdS}^2} \rp = \{5,6,10,11\}\,,
\ee
are integers as well \cite{Conlon:2021cjk, Apers:2022zjx, Apers:2022tfm}. This fascinating feature of this family of AdS vacua currently awaits an explanation and we check below in our other families whether the same is true or not.

Given that we want to compare our infinite families with the AdS distance conjecture, it is important to determine the mass scale of a tower of states that becomes light in the large flux limit. In the dual DGKT construction \cite{DeWolfe:2005uu} the large flux limit corresponds to a large volume limit and the KK scale sets the scale of a tower with a mass scale that goes to zero as the flux quanta go to infinity. Using mirror symmetry, as further discussed in appendix \ref{app:dualIIA}, we can determine the dual mass scale of a tower that becomes light in this large flux limit\footnote{By mirror symmetry the large volume limit becomes a large complex structure limit in which winding modes should become light and lead to this tower of states.}
\begin{equation}
m_{\text{tower}}^{2}\sim\frac{1}{\text{Im}(U)\text{Im}(S)^{2}}\sim\frac{(f^{0})^{3}}{(f_{1}f^{0}-(f^{1})^{2})^{\frac72}}\sim\frac{1}{f_1^{\frac72}}\,.
\end{equation}
As we discuss below, the AdS distance conjecture \cite{Lust:2019zwm}, constrains the parameter $\alpha$ that relates the mass scale of the tower to the cosmological constant via $m_{\text{tower}} \sim |\Lambda|^\alpha$. In this solution we have $\alpha =7/18$ since
\be
m_{\text{tower}} \sim \frac{1}{f_1^{\frac74}} \sim |V_{AdS}|^{\frac{7}{18}}\,.
\ee

\subsubsection{ SUSY families with $\mathbf{\alpha=1/2}$}
Next we discuss another infinite family of AdS vacua that is also parametrically controlled but not dual to the DGKT model since we have two $H_3$ flux quanta turned on. In particular, we fix the following fluxes 
\be
f^{0} = 0\,,\qquad f_{1} = 0\,, \qquad h^{0} = -3\,,\qquad h^{1} = 0\,,\qquad  h_{0} = 0 \,.
\ee
The tadpole condition in equation \eqref{eq:tadpole} is satisfied if we set $f_{0} = 4 - h_{1}f^{1}$ and we are left with two free flux parameters $h_1,f^1 \in\mathbb{Z}$. In this solution the real parts of $S$ and $U$ are equal to zero and the imaginary parts are given by 

\ba
\text{Im}(U) &=& \frac{\sqrt{\frac{9 f^{1} h_{1} +2 (-9+\sqrt{81+24   f^{1} h_{1} (-4+f^{1}  h_{1})})}{f^{1}}}}{\sqrt{15}}\,,\cr
\text{Im}(S) &=&\left[\frac{\left(-16    f^{1}  h_{1}+3
\left(9+\sqrt{81+24   f^{1} h_{1} (-4+f^{1} h_{1})}\right)\right) }{2 h_{1}^{2}}\right] \text{Im}(U)\,.
\ea

In the limit $f^{1}\to \infty$ (and for negative $h_1<0$) we find the following scaling of the moduli
\ba
\text{Im}(U) &\approx& \frac{\sqrt{(9 - 4 \sqrt{6}) h_{1}}}{\sqrt{15}}\,,\cr
\text{Im}(S) &\approx& \frac{\sqrt{6 + 8 \sqrt{\frac{2}{3}}} f^{1}}{\sqrt{-h_{1}}}\,.
\ea
So, we have parametric control since we can go to parametrically small string coupling. We can in principle also make the complex structure modulus large by an appropriate choice of $h_1$, however, this is not necessary since the Landau-Ginzburg model already takes all $\alpha'$ corrections into account \cite{Becker:2006ks}.

In the above limit of very large $f^1$ the value of the potential at the minimum is given by
\begin{equation}
    V_{AdS} \approx -\frac{27 (-h_{1})^{\frac52}}{32 \sqrt{1329 + 544 \sqrt{6}} }\frac{1}{(f^{1})^{2}}\,.
\end{equation}
Comparing the mass of the light tower from equation \eqref{eq:towermass} with the value of the scalar potential in this limit, we find that $m_{\text{tower}} \sim |V_{AdS}|^{\frac{1}{2}}$, i.e. $\alpha= 1/2$. 

In the limit where $f^{1} \rightarrow \infty$ the masses squared are,
\ba
m_{1 \pm}\hspace{0.01 cm}^{2} &=& \frac{2}{9} \lp 17 + \sqrt{6} \pm \sqrt{127 + 46 \sqrt{6}} \rp|V_{AdS}|\, ,\cr
m_{2 \pm}\hspace{0.01 cm}^{2} &=& \frac{1}{9} \lp 25 - 2 \sqrt{6} \pm \sqrt{337 + 68 \sqrt{6}} \rp |V_{AdS}|\,.
\ea
The smallest of these masses squared, $m_{2 -}^{2} = \frac{1}{9} (25 - 2 \sqrt{6} - \sqrt{337 + 68 \sqrt{6}}) |V_{AdS}| \approx - 0.260 |V_{AdS}|$, is above the Breitenlohner-Freedman bound $m_{BF}^2=-\tfrac34|V_{AdS}|$ \cite{Breitenlohner:1982jf}, as required by supersymmetry. Obviously, none of these masses are integers in AdS units and the same is true for the dual conformal scaling dimensions. Since we kept $h_1$ finite in this example, the complex structure remains finite and therefore the mirror dual type IIA families should have likewise a fixed finite volume, which might (or might not) be related to the absence of integer conformal scaling dimensions.

\subsubsection{Non-supersymmetric AdS vacua}
Lastly, we discuss here a single non-supersymmetric family of AdS vacua. We have not performed an all encompassing search for such solutions but given that they exist in the type IIA models of DGKT and are related to the supersymmetric solutions by simple sign flips of $F_4$ flux quanta, they have to exist here as well. We found one such family that is related to the supersymmetric AdS solution discussed above in subsection \ref{ssec:DGKTAdS}, by setting $f^{1}=f_{0}=0$ and flipping the sign of $f_1$.

Concretely, for $h^{0}=h^{1}=h_{1}=f^{1}=f_{0}=0$, and $f^0$ essentially fixed by the tadpole as in equation \eqref{eq:tadpoleDGKTAdS} above, we find a one parameter family of non-SUSY AdS vacua parameterized by $f_1$. The real parts of the two moduli vanish in this family, Re$(S)=$Re$(U)= 0$ and the imaginary parts are given by  
\be
\text{Im}(U) = \sqrt{\frac{5}{3}}\sqrt{-\frac{f_{1}}{f^{0}}}\,,\qquad\qquad \text{Im}(S) = \frac{2}{9} \sqrt{\frac{5}{3}}(- f_{1})^{\frac32} \sqrt{f^{0}} \,.
\ee
So, we see that both grow in the limit $f_1 \to - \infty$ and we have parametric control over these non-supersymmetric solutions. The scalar potential is given by
\begin{equation}
    V_{AdS} = - \sqrt{\frac{3}{5}} \frac{19683}{3200 (f^{0})^{\frac32} (-f_{1})^{\frac92}}\,.
\end{equation}
Since the moduli and the cosmological constant scale as for the supersymmetric counter part in subsection \ref{ssec:DGKTAdS} above, one again finds $\alpha=7/18$. 

The four masses squared for these solutions are given by
\begin{equation}
m^{2} =\left\{\frac{70}{3}, \frac{40}{3}, 6, -\frac{2}{3}\right\} |V_{AdS}|\,.
\end{equation}
The smallest of these masses squared, $m^{2} =-\frac23|V_{AdS}|$, is above the Breitenlohner-Freedman bound $m_{BF}^2=-\tfrac34|V_{AdS}|$ \cite{Breitenlohner:1982jf} and this solutions is stable, although in this case this is not guaranteed by supersymmetry.

We note that the masses squared above again give rise to dual conformal dimensions $\Delta=\{10,8,6,2 \text{ or } 1\}$ that are all integers. This was previously noticed for non-supersymmetric DGKT solutions in \cite{Conlon:2021cjk, Apers:2022zjx} and it would be interesting to extend the general analysis of \cite{Apers:2022tfm} to non-supersymmetric AdS vacua.

\subsection{AdS vacua with a large number of D3 branes}\label{ssec:AdSwithD3}
Given the fact that supersymmetric fluxes in this setup can contribute to the tadpole condition in the same way as O3 planes, we do not necessarily need the latter, however, we will keep them in the models below. We can furthermore ask whether we can find infinite families of supersymmetric vacua where a flux contribution in the tadpole can cancel an arbitrarily large number of D3 branes. This is indeed the case and we will present below two exemplary families where $N_{\text{flux}} \to -\infty$, $N_{D3} \to \infty$ while the tadpole $N_{\text{flux}} + N_{D3}=N_{O3}/2=12$ is satisfied. To the best of our knowledge such types of solution have never been discussed in the flux compactification literature before. We will present them below and then discuss potential problems and detailed features of these solutions in more detail below in subsection \ref{ssec:AdSswampland}.

\subsubsection{An infinite family with $\mathbf{\alpha = 1/2}$ and $\mathbf{N_{D3}\to\infty}$}
We will set the following four fluxes to zero $f^{1} =f_{0}= h^{0} = h_{1} =0$. Then we solve the SUSY equations $D_SW=D_UW=0$. We find supersymmetric AdS solutions with Re$(S)=$Re$(U)$=0 and the imaginary parts are stabilized at
\ba
\text{Im}(U) &=& \sqrt{\frac{-3  f^{0}h_{0}-9 f_{1} h^{1}+\sqrt{9 (f^{0} h_{0})^{2}+74 f_{1} f^{0} h_{0} h^{1}+81 (f_{1} h^{1})^{2}}}{2f^{0} h^{1}}} \,,\\
\text{Im}(S) &=& \left( \frac{-3 f^{0} h_{0} + 9 f_{1} h^{1} + \sqrt{9 (f^{0} h_{0})^{2}+74 f_{1} f^{0} h_{0} h^{1}+81 (f_{1} h^{1})^{2}}}{8 h_{0} h^{1}} \right) \text{Im}(U)\,.\nonumber
\ea
The tadpole equation \eqref{eq:tadpole} in this case reduces to 
\be
- 3 h^{1} f_{1} + h_{0} f^{0} + N_{D3} = 12\,.
\ee
Keeping $h_0$ and $f^0$ fixed and choosing a positive $h^1$, we can send $f_{1} \rightarrow \infty$. This gives rise to an infinite family of solution that requires an ever growing number of D3 branes to be present, with $N_{D3} \propto f_1$. For simplicity we study the particular example $h^1=f^0=1$. In the $f_{1} \rightarrow \infty$ limit the moduli are approximately given by
\ba
\text{Im}(U) \approx \frac{\sqrt{5 h_{0}}}{3} \,,\qquad\quad
\text{Im}(S) \approx \frac{3 \sqrt{5} f_{1}}{4 \sqrt{h_{0}}}\,.\nonumber
\ea
Thus we are at parametrically weak coupling and we can even make Im$(U)$ very large by choosing an appropriate fixed but arbitrarily large value for $h_0$.

In the limit where $f_{1}$ goes to infinity we have:
\begin{equation}
V_{AdS} \approx -\frac{2 (h_{0})^{3/2}}{25 \sqrt{5} f_{1}^{2}}\,.
\end{equation}
In the large $f_1$ limit the mass of the light tower (in Planck units) is
\begin{equation}
m_{\text{tower}}^{2} \sim \frac{1}{\text{Im}(U)\text{Im}(S)^{2}} \approx \frac{16 \sqrt{h_0}}{15 \sqrt{5} f_{1}^2}\,,
\end{equation}
which corresponds to $\alpha = 1/2$. The masses squared in this limit are
\begin{equation}
m_{1 \pm}\hspace{0.01 cm}^{2} \approx \frac{1}{27} (41 \pm 4 \sqrt{181})|V_{AdS}|\, , \qquad
m_{2 \pm}\hspace{0.01 cm}^{2} \approx \frac{1}{27} (26 \pm \sqrt{181})|V_{AdS}|\,.
\end{equation}
 
The smallest mass squared, $m_{1 -}^{2} \approx \frac{1}{27} (41 - 4 \sqrt{181}) |V_{AdS}| \approx - 0.475 |V_{AdS}|$, is above the Breitenlohner-Freedman bound $m_{BF}^2=-\tfrac34|V_{AdS}|$ \cite{Breitenlohner:1982jf}, as required by supersymmetry.

\subsubsection{An infinite family with $\mathbf{\alpha = 3/2}$ and $\mathbf{N_{D3}\to\infty}$}
Lastly, we present an infinite family that gives rise to a different value of $\alpha$, while still requiring an ever growing number of D3 branes. We choose the following fixed flux values 
\be
f^{1} = 1,\quad f_{0} = 1,\quad h^{0} = 0,\quad h^{1} =-1,\quad h_{1} = 0,\quad h_{0} = -1,\quad f^{0} =1\,,
\ee
leaving us with $f_1$ as the free parameter. There exist then supersymmetric AdS vacua in which the moduli take on the following values
\ba
&\text{Re}(U) = 0\,, \qquad &\text{Im}(U) =\frac{\sqrt{-3 - 9 f_{1} + \sqrt{9 + f_{1} (74 + 81 f_{1})}}}{\sqrt{2}}\,,\\
&\text{Re}(S) = -1\,, \qquad  &\text{Im}(S) = \left(\frac{3 - 9 f_{1} -\sqrt{9 + f_{1} (74 + 81 f_{1}) }}{8} \right) \text{Im(U)}\,.\nonumber
\ea
The tadpole equation \eqref{eq:tadpole} in this case reduces to 
\be
3 f_{1}  + N_{D3} = 13\,.
\ee
In the limit $f_{1} \to - \infty$ the above tadpole requires $N_{D3} \sim -3f_1 \to \infty$. The value of the scalar potential in this limit is 
\begin{equation}
V_{AdS} \approx - \frac{729 }{32768(-f_{1})^{\frac12}}\,.
\end{equation}
The moduli scale for $f_1 \to -\infty$ like 
\begin{equation}
\text{Im}(S) \approx \frac{8\sqrt{- f_{1}}}{3}\,, \qquad  \text{Im}(U) \approx 3\sqrt{- f_{1}}\,,
\end{equation}
and therefore 
\be
m_{\text{tower}}^{2}\sim\frac{1}{\text{Im}(U)\text{Im}(S)^{2}}\sim\frac{1}{(-f_{1})^{\frac32}}\,.
\end{equation}
This actually means that $m_{\text{tower}} \sim |V_{AdS}|^{\frac32}$, i.e. $\alpha = 3/2$ in this case.

In the limit where $f_{1} \rightarrow - \infty$ the masses become
\begin{equation}
m^{2} \approx \left\{6, \frac{10}{3}, \frac{22}{7}, -\frac{8}{27}\right\} |V_{AdS}|\,.
\end{equation}
The masses squared are above the Breitenlohner-Freedman bound $m_{BF}^2=-\tfrac34|V_{AdS}|$ \cite{Breitenlohner:1982jf}, as required by supersymmetry. Interestingly the first two masses squared give again rise to dual conformal scaling dimensions that are integers, while the later two give rise to fractional scaling dimensions: $\Delta=\{6,5, 11/3,8/3\}$.

\subsection{AdS vacua and the swampland}\label{ssec:AdSswampland}
Many explicit and widely studied constructions of AdS vacua in string theory exhibit the following two features: First, there are usually some light fields whose masses are comparable (or smaller) than the AdS scale $M_{AdS}=1/R_{AdS}=\sqrt{|V_{AdS}|/3}$ and this was conjectured to be true in all string compactifications in \cite{Gautason:2018gln}. Second, the most widely studied AdS vacua in string theory are of Freund-Rubin type \cite{Freund:1980xh, Maldacena:1997re} or exhibit similar features, by which we mean that the size of the internal space $R_{KK}$ is not parametrically smaller than $R_{AdS}$. This property was recently studied for example in \cite{Farakos:2020phe, DeLuca:2021mcj, Cribiori:2021djm,  Collins:2022nux, Tsimpis:2022orc, Apers:2022zjx, Emelin:2022cac, Cribiori:2022trc} and has led to the AdS distance conjecture \cite{Lust:2019zwm} that states that for infinite families of AdS vacua with $V_{AdS} \to 0$, there exist a tower of massive states with masses that satisfy $m_{\text{tower}} \sim |V_{AdS}|^\alpha$ for some positive $\alpha$ of order one. The strong version of this conjecture says that for supersymmetric AdS vacua $\alpha=1/2$. This conjecture has been refined in \cite{Blumenhagen:2019vgj, Buratti:2020kda}. Lastly, it was conjectured that no stable AdS vacua exist at all \cite{Ooguri:2016pdq} and all these conjectures have been used to derive important implications for the standard model of particle physics \cite{IMV1,HS,2toro,Gonzalo, Gonzalo:2021zsp,Gonzalo:2021fma}.

Against the backdrop of the above results, let us start by examining our infinite families of AdS vacua. First, let us note that in all the above families of solution the masses of the light fields $S$ and $U$ are always of the same order as $\sqrt{|V_{AdS}|}$. This means that they are all consistent with the AdS/moduli conjecture proposed in \cite{Gautason:2018gln}. 

Let us now look at the $N_{\text{flux}}=12$ solutions, which do not require the presence of D3 branes and that were discussed above in subsection \ref{ssec:AdSnoD3}. The supersymmetric AdS solutions with $\alpha=7/18$ violate the strong version of the AdS distance conjecture. A refined version of the conjecture was proposed in \cite{Buratti:2020kda} where a 4d discrete $\mathbb{Z}_k$ 3-form gauge symmetry was identified in the DGKT model and the following refined conjecture was proposed: $m_{\text{tower}} \sim \sqrt{k |V_{AdS}|}$. Given that our family of solutions is mirror dual to the DGKT AdS vacua we have a discrete $\mathbb{Z}_{f_1}$ symmetry and our solutions indeed satisfy $m_{\text{tower}} \sim \sqrt{f_1 |V_{AdS}|}$.\footnote{One could in principle work this out explicitly following \cite{Buratti:2020kda}: A 3-form gauge field with $U(1)$ gauge group arises from $F_7=dC_6$ wrapping an internal 3-cycle. This 3-form gauge field couples to the $F_3$ flux component $f_1$ and the complex structure axion Re$(U)$, which leads to the breaking of the symmetry to $\mathbb{Z}_{f_1}$. However, given the non-geometric nature of our compactifications things are more involved and it is easiest to simply rely on mirror symmetry.} 

The next family of supersymmetric AdS vacua that we discuss above satisfies the strong version of the AdS distance conjecture since it has $\alpha=1/2$. This absence of scale separation was also discovered in related IIA models in \cite{Font:2019uva}. 

This leaves us with a non-supersymmetric family of AdS solutions that is also dual to DGKT and that has $\alpha=7/18$. This is again consistent with the refined AdS distance conjecture due to the presence of a discrete symmetry that is unaffected by a simple sign flip of a flux quanta. Since these solutions are non-supersymmetric they are predicted to decay perturbatively or non-perturbatively \cite{Ooguri:2016pdq}. Given that we find that the masses squared of $S$ and $U$ are above the Breitenlohner-Freedman bound \cite{Breitenlohner:1982jf}, it is not clear whether there is a perturbative instability. Studying all possible non-perturbative decay channels or trying to identify one explicit non-perturbative decay channel is a daunting task, so we restrict ourselves here to referring to a related study of non-supersymmetric AdS vacua in the dual DGKT model \cite{Narayan:2010em}.

Finally, let us discuss the most interesting families of supersymmetric 4d $\mathcal{N}=1$ AdS vacua, namely the new families that allow for the inclusion of an arbitrarily large number of D3 branes and that are discussed in subsection \ref{ssec:AdSwithD3}. While the first one has $\alpha=1/2$ and is therefore consistent with the strong version of the AdS distance conjecture, the second one has $\alpha=3/2$, which means that the light tower is becoming light much more quickly. These solution can be made consistent with the strong version of the AdS distance conjecture by demanding $\alpha\geq 1/2$, as is already discussed in the original paper \cite{Lust:2019zwm}. Nevertheless, given that these vacua with $\alpha=3/2$ are different from all the other solutions which had $\alpha=1/2$ or smaller, they are interesting and deserve further study.

Since the later two families of supersymmetric AdS vacua have an ever increasing number of D3 branes one should worry about what that means exactly. In geometric compactifications we would expect an ever growing number of light open string modes associated with these $N_{D3}$ branes. Concretely, for $N_{D3}$ branes at separate locations the number of light open string degrees of freedom should grow like $N_{D3}$. If there is a an actual potential being generated for the D3 brane position moduli, then it seems likely that they all settle into the minimum.\footnote{At least in a geometric compactification the moduli space is compact so there are no runaway directions and for a non-trivial potential there has to exist a global minimum. Any potential that is generated for the D3 brane position moduli should be small in our limit of parametrically weak string coupling, so these position moduli should be light.} We can of course also always choose to place all the $N_{D3}$ on top of each other and since they are mutually BPS there should be no force between them. This would then lead to a number of light degrees of freedom that grows even faster like $N_{D3}^2$. Due to the species bound \cite{Veneziano:2001ah, Arkani-Hamed:2005zuc, Dvali:2007hz, Dvali:2007wp}, this leads to a UV cutoff that goes like $\Lambda_{UV} \sim M_{pl}/\sqrt{N_{D3}^2} = M_{pl}/N_{D3}$. In our first family of AdS vacua one finds that $\Lambda_{UV} \sim 1/f_1 \sim m_{\text{tower}}$. So, the UV cutoff from the species bound scales in the same way as the infinite tower of light modes. In the second example with $\alpha=3/2$ one finds that $\Lambda_{UV} \sim 1/f_1 \sim m_{\text{tower}}^{\frac43}$. This means that the species bound is even lower than the tower of light states that comes down rather quickly in this case anyways. Note that the previous discussion is based on the geometric intuition that might well carry over to these non-geometric setups. However, the actual open string spectrum for D3 branes in these model was not worked out in the previous literature. We leave it as an interesting task for the future to check the light open string degrees of freedoms in these models.

Slightly disconnected from the different AdS conjectures discussed above, we lastly would like to point out the most interesting and most surprising feature of these AdS solutions with $N_{D3} \to \infty$: The fluctuations along the AdS$_4$ directions of the open string modes on these branes should give rise to gauge groups with arbitrarily large rank! For example, if we place all $N_{D3}$ branes on top of each other one would naively expect an $SU(N_{D3}/2)$ gauge group.\footnote{The tadpole condition in equation \eqref{eq:tadpole} counts D3 branes in the covering space, hence there can be at most $\lfloor N_{D3}/2\rfloor$ freely moving D3 branes in the quotient space. If $N_{D3}$ is odd then one D3 brane would necessarily be stuck on top of an O3 plane.} String universality in higher dimensions with higher amount of supersymmetry leads to fairly low ranks for the gauge group, which seems in stark contrast with the solutions above. This is a by now very active area of research following the initial work of \cite{Adams:2010zy, Kim:2019vuc, Kim:2019ths, Cvetic:2020kuw, Montero:2020icj}. Currently, there is no argument in the literature that forbids 4d $\mathcal{N}=1$ (not scale separated) AdS solutions with an $SU(N)$ gauge group for arbitrarily large $N$. So, it seems in principle possible that such solutions do exist in the barely explored part of the string landscape that we study here. We again add as a word of caution that the open string spectrum for these D3 branes has not been worked out and therefore it could hypothetically not contain any massless open strings or no gauge fields at all. It would be very interesting to check this explicitly and we hope to do this in the future.

\section{de Sitter vacua}\label{sec:dS}
Lastly, we would like to comment on the existence of dS vacua in this setup. Given that the K\"ahler potential can receive string loop corrections, one finds that non-supersymmetric solutions can cease to exist, if corrections are large. Thus, unless they are at weak coupling one should not trust non-supersymmetric solutions. All dS solutions in the models discussed here will have a string coupling that is not that much smaller than 1 and it is therefore not clear whether they can be trusted. Nevertheless, we discuss them for the following two reasons: Firstly, they were recently studied in \cite{Ishiguro:2021csu} and we would like to comment on and extend these previous results. Secondly, dS vacua are notoriously difficult to construct in purely classical scalar potentials \cite{Andriot:2019wrs} and only very few explicit solutions without tachyons exist in the literature \cite{Kallosh:2018nrk, Blaback:2018hdo, Andriot:2021rdy, Shukla:2022srx}. Therefore, it is interesting to check whether they also exist in our simple models or not.

Unstable dS solutions, i.e. solutions with a tachyonic direction and the correct tadpole for the $1^9$ model, $N_{\text{flux}}=12$, were found in \cite{Ishiguro:2021csu}. Interestingly, the authors of that paper performed a scan over flux values that do not satisfy the tadpole condition and they found that stable dS vacua exist for a large $N_{\text{flux}} \sim \mathcal{O}(100)$. They also noticed that the ratio of stable dS vacua to all randomly generated vacua grows with $N_{\text{flux}}$ (see figure 9 in \cite{Ishiguro:2021csu}). The smallest $N_{\text{flux}}$ value that was giving rise to a stable dS solution in figure 9 in \cite{Ishiguro:2021csu} is larger than 66 and the smallest, explicitly listed, stable dS solution in table 5 of that paper has $N_{\text{flux}}=74$. While this is substantially larger than the allowed $N_{\text{flux}}=12$ in the $1^9$ model, it is not that much larger than the allowed $N_{\text{flux}}=40$ in the $2^6$ model.

\subsection{Explicit dS solutions}
An explicit tachyonic dS extremum with $N_{\text{flux}}=12$ was previously found in \cite{Ishiguro:2021csu}. The corresponding fluxes are 
\be
f^{0} = 4, \quad f^{1} = 8,\quad f_{1} = 7,\quad f_{0} = -17,\quad h^{0} =1,\quad h^{1} = 1,\quad h_{1} = 1\,, \quad h_0=-2\,.
\ee
Given that $N_{\text{flux}}=12$ this is a solution to the $1^9$ model which does not require D3 branes since the fluxes cancel the negative O3 plane charge. The moduli are stabilized at the following values
\ba
&\text{Re}(U) \approx 0.544 \,,\qquad\qquad &\text{Im}(U) \approx 1.11 \,,\cr
&\text{Re}(S) \approx 7.72 \,, \qquad \qquad &\text{Im}(S) \approx 5.19 \,.
\ea
The value of the scalar potential is given by $V_{dS}\approx 1.72 \times 10^{-4}$. The masses squared for the four real scalar fields in the unstable dS extremum are given by
\be
m_1^2 \approx 0.0226\,,\qquad m_2^2 \approx 0.0157\,,\qquad m_3^2 \approx  0.00143 \,,\qquad m_4^2 \approx -0.00119 \,.
\ee

So, there are unstable dS solutions like the one above and, as mentioned previously, there are also metastable dS vacua, if one ignores the tadpole and lets $N_{\text{flux}}$ become fairly large. Therefore, one should ask what the lowest possible value for $N_{\text{flux}}$ is that still gives rise to metastable dS solutions. We have not been able to answer this question in full generality. However, we noticed that unstable and metastable dS solutions still exist when we set four fluxes to zero: $f^{1} = f_{0} = h^{0} = h_{1} = 0$. We then studied the full parameter space spanned by the remaining four fluxes, while ignoring the tadpole. This led us to discover infinite families of solutions that transition from AdS to unstable dS and then to metastable dS, if we vary the fluxes. Within these family we identified the smallest possible $N_{\text{flux}}$ that has integer quantized fluxes and gives rise to metastable dS solutions. We find that the only possible value below $N_{O3}/2=40$ for the $2^6$ model is $N_{\text{flux}}=30$.\footnote{The next larger values of $N_{\text{flux}}$ that give rise to metastable dS solutions in our restricted model with only four non-zero fluxes are $N_{\text{flux}}=\{59,60,61\}$. This is too large to be compatible with the tadpole cancellation condition.} For this value there are four different metastable dS solutions. Three have Im$(S)<1$ and are therefore expected to receive substantial string loop corrections. The fourth one with the fluxes
\be
f^{0} = 33, \quad f^{1} = 0,\quad f_{1} = -1,\quad f_{0} = 0,\quad h^{0} =0,\quad h^{1} = -1,\quad h_{1} = 0\,, \quad h_0=1\,,
\ee
has a metastable dS vacuum at
\ba
&\text{Re}(U) = 0 \,,\qquad\qquad &\text{Im}(U) \approx 0.299 \,,\cr
&\text{Re}(S) = 0 \,, \qquad \qquad &\text{Im}(S) \approx 1.32 \,.
\ea
The value of the scalar potential is given by $V_{dS} \approx 0.00524$. The masses squared for the four real scalar fields in the dS minimum are given by
\be
m_1^2 \approx 3.31\,,\qquad m_2^2 \approx 1.29\,,\qquad m_3^2 \approx  .302 \,,\qquad m_4^2 \approx 0.0999 \,.
\ee
Given that $N_{\text{flux}}=30$ this is a solution to the $2^6$ model which does require $N_{D3}=10$ D3 branes. Thus, there should be additional light open string moduli associated with those D3 branes. 

It would be interesting to extend our full analysis beyond the restriction $f^{1} = f_{0} = h^{0} = h_{1} = 0$ and check whether there exist metastable dS solutions in these models that are at smaller string coupling and/or that do not require D3 branes in order to satisfy the tadpole. Due to the mirror symmetry that relates our above models to models with $H_3$ flux and non-geometric $Q$ flux there should be also a connection to the metastable dS solution found in 2009 in \cite{deCarlos:2009qm}. Note however, that the latter also required geometric and/or non-geometric fluxes in the type IIB duality frame since $h^{1,1} \neq 0$ and thus they are less controlled then the models we discussed in this paper due to the risk of large $\alpha'$ corrections.

\subsection{dS extrema and the swampland}
The very existence of dS vacua in string theory was first questioned in \cite{Danielsson:2018ztv, Obied:2018sgi} and a variety of refined dS swampland conjectures were proposed in 2018 in for example \cite{Andriot:2018wzk, Garg:2018reu, Hebecker:2018vxz, Ooguri:2018wrx, Andriot:2018mav}. All of these conjectures forbid metastable dS solutions. However, given that our metastable dS solution above is expected to receive substantial string loop corrections, it does not invalidate these conjectures. The previously discovered unstable dS extremum of \cite{Ishiguro:2021csu} has $e^{\phi} \approx .5$ and does not require D3 branes. It is thus in much better shape, however, given that it is unstable with large $|\eta| \approx 7$ it is not really incompatible with any of the refined dS swampland conjectures.

It would be interesting to study this simplified model or related more complicated setups to see whether one can find metastable dS vacua at weak coupling and without D3 branes. While there is no obstruction to this, it was recently shown in the context of type IIA flux compactifications that dS solutions cannot exist in a parametrically controlled region \cite{Junghans:2018gdb, Banlaki:2018ayh}. While these papers mostly focused on geometric type IIA flux compactifications they also discuss more exotic ingredients like non-geometric fluxes which makes them applicable to all the type IIA flux compactifications that are the mirror dual of our type IIB setup. Thus, they actually apply also to our non-geometric type IIB models. This means there should be no parametrically controlled dS solutions, i.e. no solutions with a free flux parameter that we can send to infinity to get Im$(S) \to \infty$. However, there is no obvious reason why well-controlled dS solutions with Im$(S) \gg 1$ cannot exist in the setup discussed in this paper.

\section{Conclusion}\label{sec:conclusion}
In this paper we have studied type IIB flux compactifications based on Landau-Ginzburg orientifolds. We have focused on models that are non-geometric in the sense that $h^{1,1}=0$, i.e., there are no K\"ahler moduli. This barely studied class of models was originally introduced in \cite{Becker:2006ks, Becker:2007dn} and allows for full moduli stabilization. We have revisited these models and discovered a variety of interesting new families of solutions. We have contrasted these solutions with several swampland conjectures (see \cite{Ishiguro:2021csu} for recent related work). 

Concretely, we have explored the four dimensional landscape of two models which are mirror duals to type IIA string theory on rigid Calabi-Yau orientifolds, i.e., Calabi-Yau manifolds with $h^{2,1}=0$. After including $H_3$ and $F_3$ fluxes our models are dual to type IIA flux compactifications with both metric and non-geometric fluxes, so our analysis goes beyond (and includes) setups such as DGKT \cite{DeWolfe:2005uu}. However, while non-geometric fluxes are not under control in type IIA supergravity models, we have only regular (and well understood) $H_3$ and $F_3$ fluxes in the mirror dual Landau-Ginzburg models in IIB. Furthermore, there exists a very powerful non-renormalization theorem that protects the superpotential from receiving any corrections at all \cite{Becker:2006ks}. For simplicity we have focused here on an isotropic two moduli ($SU$) model, which is not the most general setup, but it is already enough to provide us with new interesting results that we now sum up. 

In this work we have provided additional arguments which point to the existence of fully stabilized 4d $\mathcal{N}=1$ Minkowski vacua. While these were originally discovered in \cite{Becker:2006ks, Becker:2007dn}, we managed to find infinite families of Minkowski vacua and we have argued that they are in principle compatible with existing swampland conjectures like \cite{Gautason:2018gln, Palti:2020qlc}. Furthermore, we have proven that although the masses do receive corrections, they can never become zero and there cannot arise any flat directions even when including all unknown corrections.

We have also found several new infinite families of AdS vacua, which are not connected to the aforementioned family of Minkowski vacua. By taking some particular flux combinations to infinity (often simply one of the fluxes) these AdS solutions approach Minkowski space. However, in every example we have argued using mirror symmetry that there is a tower of states becoming light with a certain power $\alpha \geq 1/2$ of the cosmological constant, i.e. $m_{\text{tower}} \sim |V_{AdS}|^{\alpha}$. Thus, our results in this regard are consistent with the AdS distance conjecture \cite{Lust:2019zwm}. However, since our models are essentially a generalization of the DGKT models in type IIA, we also identified supersymmetric and non-supersymmetric infinite families of AdS vacua in a subset of our model, which have $\alpha=7/18$ like the original examples in DGKT \cite{DeWolfe:2005uu}. For similar reasons as discussed in \cite{Buratti:2020kda} we find agreement of these families with the refined version of the AdS distance conjecture due to a large discrete 3-form gauge symmetry. For the nonsupersymmetric infinite family of AdS vacua, our moduli $S$ and $U$ acquire masses squared above the BF bound. These vacua arise in a regime of parametric control but should be unstable according to the conjecture in \cite{Ooguri:2016pdq}. It would be interesting to analyze possible decay modes for our family of non-supersymmetric AdS solutions.

As explained in \cite{Becker:2007dn}, due to the non-geometric nature of our models, the K\"ahler potential acquires an unfamiliar factor of 4 whose main effect is to allow supersymmetric fluxes that are not imaginary self-dual. This actually allows the $H_3$ and $F_3$ fluxes to contribute to the D3/O3 tadpole condition with either sign. Interestingly, this enables us to construct new infinite families of supersymmetric AdS$_4$ vacua with an unbounded number of spacetime filling D3 branes. This is possible because the flux contribution to the tadpole can have the same sign as that of O3 orientifold planes and we can make it arbitrarily large. This arbitrarily large flux then requires an arbitrarily large number of D3 branes. Given that all these solutions are consistent with the AdS distance conjecture there is an infinite tower of massive states becoming light when we increase the flux and the number of D3 branes at the same time. Furthermore, it is expected that there are large numbers of massless open string modes that are associated with those D3 branes, leading to an ever decreasing species bound $\Lambda_{UV} \sim M_{pl}/N_{D3}$. Nevertheless, it seems naively possible to get a very large rank for the gauge group in these 4d $\mathcal{N}=1$ AdS vacua. It would be very interesting to study this further and see whether these solutions are indeed trustworthy or suffer from some inconsistencies. 

Finally, we have been able to find a metastable de Sitter vacuum that requires some number of D3 branes to be present. However, this vacuum does not arise at weak coupling and there is no argument preventing perturbative and non-perturbative corrections from destroying it.

Given the current large amount of activity in the swampland program, it is very important to keep exploring all different areas of the string landscape, in particular, areas that are truly stringy in the sense that they do not have a geometric supergravity description. In this paper we have revisited and extended previous studies of type IIB flux compactifications in the absence of K\"ahler moduli, i.e. for $h^{1,1}=0$. We found several intriguing results which could be natural in this rather unexplored corner of the string landscape and that deserve further study in the future. 

\section*{Acknowledgments}
We like to thank Thomas Grimm, Michael Haack, Miguel Montero, Jakob Moritz and Thomas Van Riet for useful discussions. This work is supported in part by the NSF grant PHY-2013988.

\appendix

\section{Details of the dual type IIA models}\label{app:dualIIA}

\subsection{K\"ahler potential}\label{sapp:K}
As mentioned in the text the way in which one derives the formula for the K\"ahler potential is by using mirror symmetry. The usual formula for the K\"ahler potential in Type IIA flux compactifications on Calabi-Yau manifolds is \cite{Grimm:2004ua}
\begin{equation}
K=-\text{log}\left[\frac{4}{3}\int J\w J\w J\right]-2\text{log}\left[2\int\text{Re}\left(C\Omega_{3}\right)\wedge\star\text{Re}\left(C\Omega_{3}\right)\right]\,,
\end{equation}
where $J$ is the K\"ahler form and $\Omega$ the holomorphic 3-form. The volume is given by $\text{vol}_{6} =\frac16 \int J\w J\w J$. The so-called 4d dilaton is defined via  $e^{D}=e^{\phi}/\sqrt{\text{vol}_{6}}$ and  $C\overline{C}\int\Omega\wedge\overline{\Omega}=e^{-2D}$. The supergravity fields
are introduced by expanding the complexified K\"ahler form and the complexified holomorphic 3-form \cite{Grimm:2004ua}
\ba
J_{c}&=&B_{2}+\rmi J= \sum_{a=1}^{h^{(1,1)}}T^{a}\omega_{a}\,,\\
\Omega_{c}&=&C_{3}+2\rmi\text{Re}\left(C\Omega_{3}\right)= S \alpha_{0} + \sum_{k=1}^{h^{(2,1)}}U^{k}\alpha_{k}\,.
\ea

When $h^{(2,1)}=0$ there are no complex structure moduli. We can always write the volume in terms of the triple intersection number $\kappa_{abc}=\int \omega_a \w \omega_b \w \omega_c$ of the Calabi-Yau manifold, which leads (up to a constant) to the K\"ahler potential
\begin{equation}
K=-\text{log}\left[\frac{\rmi}{6}\kappa_{abc} \left(T^{a}-\overline{T}^{a}\right)\left(T^{b}-\overline{T}^{b}\right)\left(T^{c}-\overline{T}^{c}\right)\right]-4\text{log}\left[-\frac{\rmi}{2\sqrt{2}}\left(S-\overline{S}\right)\right]\,.
\end{equation}

Mirror symmetry simply exchanges the $h^{1,1}$ K\"ahler moduli $T^a$ with the $h^{2,1}$ complex structure moduli $U^k$. Since we have no complex structure moduli the mirror dual K\"ahler potential is the one given above in equation \eqref{eq:K}, if one restricts to the torus bulk moduli and sets them all equal \cite{Becker:2007dn}. The superpotential can be derived in the same way but was also argued for directly in type IIB in \cite{Becker:2006ks}.

\subsection{KK towers}\label{sapp:tower}
In this section, following the original work \cite{DeWolfe:2005uu}, we quickly review how to derive the KK scale in type IIA flux compactifications. Using mirror symmetry we can then derive the mass scale for a light tower in the non-geometric type IIB flux compactifications discussed in this paper. As on the type IIA side, this is not proven to be always the lightest tower but no other lighter tower is expected to arise in the type IIA side, so presumably the same is true on the type IIB side. Also, our infinite families of AdS vacua are all consistent with the refined AdS distance conjecture \cite{Lust:2019zwm, Buratti:2020kda}, which means that this is likely the relevant tower of massive states.

The KK scale in type IIA flux compactifications is controlled by the internal volumes of 2-cycles, Im$(T^a)$. In the isotropic limit where we set the three bulk 2-cycles of the torus equal we will simply use Im$(T)$ to describe this volume. So, we know that $m_{KK}^2$ scales like $1/\text{Im}(T)$. Compactifying from 10d to 4d and then going to 4d Einstein frame introduces an extra factor and the correct KK scale is given by
\begin{equation}
m_{KK}^{2} \sim \frac{1}{\text{vol}_6 e^{-2\phi} \text{Im}(T)}=\frac{1}{(\text{Im}(S))^2 \text{Im}(T)}\,.
\end{equation}
Again using mirror symmetry, we find a dual massive tower with masses that scale like
\begin{equation}\label{eq:towermass}
m_{\text{tower}}^{2} \sim \frac{1}{\left(\text{Im}S\right)^{2}\text{Im}(U)}\,.
\end{equation}

\bibliographystyle{JHEP}
\bibliography{refs}

\end{document}